\documentclass{aa}
\usepackage{psfig}

\begin{document}

\def\lta{\la}
\def\gta{\ga}
\def\uu {4U~1700$+$24~}
\def\xmm  {\textit{XMM-Newton~}}
\def\cha {\textit{Chandra~}}
\def\xte {\textit{RXTE~}}
\def\mdot {\dot M}
\def\msole{~M_{\odot}}
\def\ltsima{$\; \buildrel < \over \sim \;$}
\def\lsim{\lower.5ex\hbox{\ltsima}}
\def\gtsima{$\; \buildrel > \over \sim \;$}
\def\gsim{\lower.5ex\hbox{\gtsima}}

\title{Discovery of a redshifted X-ray emission line in the symbiotic
neutron star binary \uu}

        \author{A. Tiengo\inst{1}$^{,}$\inst{2},
                D. K. Galloway\inst{3},
                T. di Salvo\inst{4},
                M. M\'endez\inst{5},
                J. M. Miller\inst{6}$^{,}$\inst{7},
                J. L. Sokoloski\inst{6}$^{,}$\inst{7},
                and M. van der Klis\inst{8}}

   \offprints{A.Tiengo, email: tiengo@mi.iasf.cnr.it}

\institute{Istituto di Astrofisica Spaziale e Fisica Cosmica, Sezione di
        Milano ``G.Occhialini''- INAF, via Bassini 15, Milano I-20133, Italy
        \and
        Universit\`{a} degli Studi di Milano,
        Dipartimento di Fisica, via Celoria 16, I-20133 Milano, Italy
        \and
        Center for Space Research,
        Massachusetts Institute of Technology, Cambridge, MA 02139-4307
        \and
        Dipartimento di Scienze Fisiche ed Astronomiche, Universit\`{a} di
        Palermo, via Archirafi 36, 90123 Palermo, Italy
        \and
        SRON, National Institute for Space Research, Sorbonnelaan 2,
        NL 3584 CA, Utrecht, the Netherlands
        \and
        Harvard-Smithsonian Center for Astrophysics, 60 Garden Street,
        Cambridge, MA 02138
        \and
        NSF Astronomy and Astrophysics Fellow
        \and
        Astronomical Institute ``Anton Pannekoek'', University of
        Amsterdam and Center for High-Energy Astrophysics,
        Kruislaan 403, NL 1098 SJ Amsterdam, the Netherlands}


\abstract{ We present the spectral analysis of an \xmm observation
of the X-ray binary \uu, performed during an outburst in August
2002. The EPIC-PN spectrum above 1 keV can be modeled by a blackbody
plus Comptonization model, as in previous observations. At lower
energies, however, we detect a prominent soft excess, which we model
with a broad Gaussian centered at $\sim 0.5$ keV. In the high
resolution RGS spectrum we detect a single emission line, centered
at 19.19$^{+0.05}_{-0.09}$ \AA. We discuss two possible
interpretations for this line: O~{\small VIII} at redshift $z=
0.012^{+0.002}_{-0.004}$ or Ne~{\small IX} at redshift $z \sim 0.4$.
\keywords{X-rays: individuals: \uu --- stars: neutron
 --- binaries: symbiotic}
}

\authorrunning{Tiengo et al.}

\maketitle

\section{Introduction}

Based on its optical identification with the M2 III giant HD154791
(Garcia et al. 1983; Masetti et al. 2002; also known as V934 Her),
the X-ray source \uu is classified as both a Low Mass X-ray Binary
(Liu et al. 2001) and a symbiotic-like binary (Garcia et al. 1983).
The lack of signatures of binarity and the initially marginal
positional coincidence (Morgan \& Garcia 2001) originally made this
identification uncertain. More recently, the detection of radial
velocity variations from the proposed optical counterpart with the
same period as X-ray brightness variations (Galloway et al. 2002)
has added support to the identification.

In X-rays, \uu has long periods in which it is faint,
and occasional episodes in which the flux increases by more than an
order of magnitude.  Assuming a distance of 420 pc (Masetti et
al. 2002), the 2--10 keV luminosity of
\uu varies from 2$\times$10$^{32}$, which is much higher than
expected from an isolated M giant of this spectral type, to 10$^{34}$
erg s$^{-1}$.
Masetti et al. (2002) reported the analysis of six observations from
five different X-ray
satellites, including an
observation with the {\it Rossi X-Ray Timing Explorer} ({\it RXTE};
Jahoda et al. 1996) taken during a $\sim 100$-day outburst in
November 1997.  Although no coherent pulsations or QPOs were
detected, their analysis
shows that the X-ray flux is also variable on short time scales
($\sim$ seconds).

In July 2002, the \textit{RXTE} All-Sky Monitor light curve of 4U
1700+24 indicated that the source was undergoing a new outburst, and
a Target of Opportunity observation was performed by the \xmm
satellite. In the following sections we discuss the analysis of the
spectroscopic X-ray data from this observation.

\section{Observations and data reduction}

\uu was observed with the \xmm satellite (Jansen et al. 2001) on
2002 August 11. The {\it XMM-Newton}\/ X-ray optics consist of three
nested Wolter-I mirror assemblies, illuminating five X-ray detectors
which operate simultaneously. CCDs sensitive to photons in the range
0.15-15 keV are situated at the focal point of each the three
mirrors, two MOS and one PN types, comprising the {\it European
Photon Imaging Camera} (EPIC; Str\"{u}der et al. 2001, Turner et al.
2001). Two of the mirrors also illuminate the {\it Reflection
Grating Spectrometers} (RGS; den Herder et al. 2001) which disperse
photons in the range 5--35~\AA\ onto a pair of dedicated off-axis
CCDs. The PN and MOS cameras have spectral resolutions of $\simeq$80
eV and $\simeq$70 eV (at 1 keV) and point spread functions (PSFs) of
$\simeq$6$''$ and $\simeq$5$''$ for the PN and MOS cameras,
respectively. The RGS spectral resolution is 0.04 \AA\, although it
has a significantly smaller effective area ($\simeq$100 cm$^2$ at 1
keV for each of the two units, compared to $\simeq$1200 cm$^2$ for
the PN and $\simeq$900 cm$^2$ for the sum of the two MOS detectors)
and cannot produce 2-dimensional images.

The exposure time of the \uu observation
(corrected for the dead--time) was 6 ks for the PN, 7 ks for the
MOS, and 8 ks for the RGS. To avoid excessive optical loading on the
PN and MOS cameras from
the bright optical companion of \uu (V=7.6), the ``thick'' filter
was selected. The X-ray brightness of \uu during this high-state
observation necessitated the use of Small Window operating mode, in
which only a small portion (about 4$'\times$4$'$ for the PN and
2$'\times$2$'$ for the MOS) of the CCD is read out in a reduced
frame time (6 ms for PN and 300 ms for MOS).  The short readout time
of this mode helps reduce errors due to photon pile--up, which
occurs when two (or more) photons hit nearby pixels
during a single CCD exposure, producing an event
that is indistinguishable from
a single, higher energy photon.
The main effects of photon pile--up are an incorrect reconstruction
of the event energies
and an underestimation of the source count rate.

All the data were processed using the {\it XMM-Newton Science
Analysis System} (SAS, version
6.1.0)\footnote{http://xmm.vilspa.esa.es/external/xmm\_sw\_cal/sas\_frame.shtml}.
The use of Small Window mode only partially mitigated the effects of
pile--up\footnote{The presence of pile--up in a dataset can be
investigated by examining the energy distribution of single, double,
and multiple events (composed, respectively, of one, two, and
greater than two adjacent pixels being above the detection
threshold)} in the PN and MOS data, and so in our extraction of
source photons we excluded the central part of the source PSF, where
the incident count rate per pixel is the highest. The EPIC spectra
were thus extracted from annular regions centered on the source,
with inner and outer radii of 10$\arcsec$ and 40$\arcsec$,
respectively. With these regions, we were able to completely remove
the portion of the image affected by pile--up from the PN data, but
not from the MOS data.  The choice of an even larger inner radius
for MOS would have greatly reduced the number of valid counts.
Therefore, the MOS spectra are not discussed in this paper. To
generate the PN spectrum, only single and double events were
retained, and the resulting spectrum was rebinned both to have at
least 30 counts in every energy channel and at most three bins per
PN energy resolution element. A background spectrum was extracted
from a region far enough from \uu to avoid contamination by source
photons. In the 0.2--12 keV range, the background contributes less
than 0.5\% of the total count rate for most of the observation,
increasing to 3\% during the last 1500 seconds.

The RGS spectra (source and background) were extracted, and response
matrices calculated using the standard reduction procedures. Since
the source photons are widely spread along the dispersion axis, the
RGS data are not affected by pile--up. The RGS  source spectra were
also rebinned in order to have at least 30 counts in every channel.

\section{Analysis and results}

\subsection{Spectral Continuum}

In order to compare the \xmm spectrum of \uu with previous
observations, we fitted the PN spectrum in the 0.3--12 keV range
with the two-component model used by Masetti et al. (2002): a
blackbody plus Comptonization (COMPST; Sunyaev \& Titarchuk 1980),
with absorption fixed to the Galactic value of $N_{H}$ =
4$\times$10$^{20}$ cm$^{-2}$ (Dickey \& Lockman 1990). The resulting
fit is unacceptable ($\chi_{red}^2$ = 4.05 for 273 degrees of
freedom) due to a large excess at low energies. The fit can be
improved by adding a component that contributes below $\simeq 1$
keV. We obtain the best results by modeling this soft excess with a
broad ($\sigma\simeq$0.1 keV) Gaussian line centered at $\simeq$0.5
keV.  We report the spectral parameters in Table 1. We show the PN
spectrum and its residuals with respect to this model in Figure 1.
Although even our final fit is only marginally acceptable
($\chi_{red}^2$ = 1.19 for 270 d.o.f., corresponding to a null
hypothesis probability of 0.018), the addition of a systematic error
of 1.5\%, which is compatible with the current calibration
uncertainties of the PN, reduces $\chi_{red}^2$ to 1. No narrow
features are detected in the PN spectrum. In particular, the
3$\sigma$ upper limit on the equivalent width of a Fe K-$\alpha$
fluorescent line at 6.4 keV is $\simeq 15$ eV.

\begin{figure}
\psfig{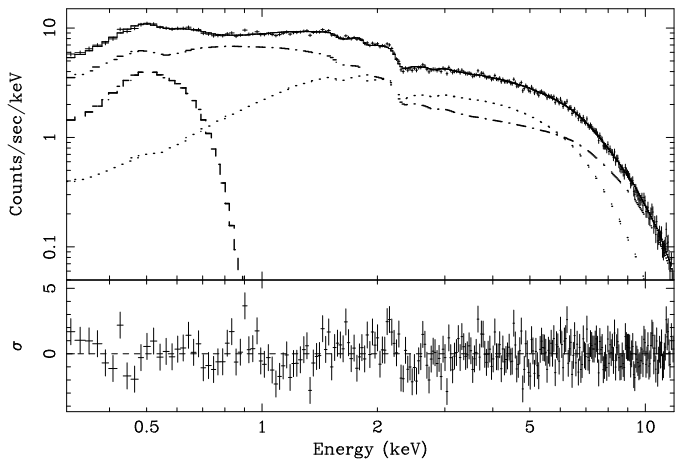} \caption{The PN count
rate spectrum of \uu, and residuals (in units of $\sigma$) with
respect to the best-fit model described in Table 1. The individual
model components are also shown: dashed line for the broad Gaussian,
dotted line for the blackbody and dash-dotted line for the
Comptonization model. \label{fig1}}
\end{figure}

\begin{table*}[htbp!]
\begin{center}
  \caption{Best fit parameters of the PN spectrum in the 0.3--12 keV energy range.
Errors are at the 3 $\sigma$ confidence level for a single
interesting parameter.  }
    \begin{tabular}[c]{cc}
\hline \hline
N$_H$ (10$^{20}$ cm$^{-2}$) & 4 (fixed) \\
kT$_{e}$ (keV)         & 4.4$^{+2.8}_{-0.9}$                   \\
$\tau$                 & 18.5$^{+1.7}_{-5.0}$           \\
N$_{C}^{(a)}$          & 0.0229$\pm0.0007$             \\
kT$_{BB}$ (keV)       & 1.40$\pm$0.03               \\
R$_{BB}$ (m)$^{(b)}$ & 113$\pm$3            \\
E$_{\rm line}$ (eV)   & 505$^{+13}_{-12}$                  \\
$\sigma$ (eV)         & 112$^{+25}_{-20}$                  \\
Equivalent width (eV) & 198$^{+34}_{-28}$                     \\
\hline
$\chi^2_{red}$ (d.o.f.)& 1.188 (270)               \\
\hline
0.3--2 keV flux$^{(c)}$& 8.4$\times$10$^{-11}$     \\
2--10 keV flux$^{(c)}$ & 5.3$\times$10$^{-10}$     \\
\hline \hline

\end{tabular}
\end{center}
\begin{small}
$^{(a)}$ Normalization of the COMPST model in XSPEC version 11.3\\
$^{(b)}$ Radius at infinity assuming a distance of 420 pc\\
$^{(c)}$ Absorbed flux in units of ergs cm$^{-2}$ s$^{-1}$\\

\end{small}
\label{spectra}
\end{table*}

\subsection{The emission line at 19.2 \AA}

The high spectral resolution of the RGS allows us to search for
narrow features in the X-ray spectrum of \uu at low energies. In
Figure 2, we show the first-order RGS spectrum of \uu and residuals
with respect to the best-fit model derived from analysis of the PN
data.  In addition to some deviations from the continuum model due
to problems in the cross-calibration between the PN and RGS
instruments (Kirsch et al.  2004), an emission line is apparent at
about 19 \AA\, (Fig. 2). The addition of a Gaussian component in the
model accounts for the residuals in this region of the spectrum (see
bottom panel of figure 2) and improves the $\chi^2$ from  1.65 (for
765 d.o.f.) to 1.54 (for 762 d.o.f.). The emission-line parameters
are: $\lambda$ = 19.19$^{+0.05}_{-0.09}$ \AA, $\sigma$ =
$3.9_{-1.3}^{+2.7}$ eV, line flux equals $(4.9^{+1.9}_{-1.6})\times
10^{-4}$ photons cm$^{-2}$ s$^{-1}$, and equivalent width equals
9.7$^{+3.8}_{-3.1}$ eV (all the quoted errors are 3$\sigma$ errors).

In the X--ray atomic lines catalogue ATOMDB
1.3.1\footnote{http://asc.harvard.edu/atomdb/}, only very low
emissivity transitions
 are found within 3$\sigma$ of the line position;
most
are satellite lines of H-like and He-like oxygen, with
a maximum emissivity of 8.4$\times 10^{-18}$ photons cm$^3$
s$^{-1}$. The Ly-$\epsilon$ line of N~{\small VII} (which is not
included in the ATOMDB database) has a wavelength compatible with the
observed emission line
($\lambda$ = 19.118 \AA, Verner et al. 1996a), but the probability
for this transition is very low. The Ly-$\alpha$ transition of
H-like oxygen (O~{\small VIII}) has an emissivity more than 2 orders
of magnitude higher than that of the lines mentioned above (which
makes it typically one of the strongest lines
found in cosmic X--ray sources) and a rest-frame wavelength of 18.97
\AA,
slightly smaller than the value we find for the emission line
shown in Figure 2.   It is therefore possible
that the X-ray emission line in \uu is
O~{\small VIII}
at redshift {\bf $z= 0.012^{+0.002}_{-0.004}$}.

The accuracy of the RGS wavelength scale is better than 10 m\AA\,
(Pollock 2004), with an additional 2.3 m\AA\, of systematic error
for each arcsecond error in the pointing or source coordinates. For
the \uu observation, the additional error is smaller than 10 m\AA,
since the EPIC positional accuracy is better than 4$''$ and the RGS
data were processed using source coordinates derived from the EPIC
images. Therefore, we exclude the possibility that the wavelength
shift from the O~{\small VIII} rest-frame position is due to an
incorrect wavelength scale in the RGS instrument.

Other smaller structures are also present in the residuals. In
particular, the data differ substantially from the model around the
instrumental oxygen edge at $\simeq 23.5$ \AA, where the RGS
effective area calibration is complex. We can improve the RGS
spectral fit slightly ( $\chi^2_{red}$=1.51 for 761 d.o.f.) by
including an overabundance of neutral oxygen in the photoelectric
absorption model (with cross section from Verner at al. 1996b).
Keeping the hydrogen column density fixed at the value of
4$\times$10$^{20}$ cm$^{-2}$, we obtain a best-fit value for the
oxygen abundance of 2.0$^{+0.3}_{-0.5}$ times the solar value
(Anders \& Grevesse 1989).  On the other hand, two instrumental
absorption features at 23.05 and 23.35 \AA\, contribute $\sim
2\times$10$^{17}$ cm$^{-2}$ to the oxygen column density (de Vries
et al. 2003) and could also explain most of the detected
overabundance of oxygen.

If the emission line at 19.19 \AA\, is redshifted O~{\small VIII},
we might expect spectral features from other ionization states of
oxygen. The signature of He-like oxygen is a triplet of X-ray lines
at 21.6, 21.8, and 22.1 \AA. Some residuals are in fact present in
the RGS spectrum around 21.8 \AA, which corresponds to either the
rest-wavelength of the intercombination line or a resonance line at
$z$=0.01. Fitting the residuals at 21.8 \AA\, with a Gaussian of the
same width as the O~{\small VIII} line candidate ($\sigma$ = 4 eV)
gives a 3$\sigma$ upper limit for the equivalent width of 5 eV.
Fully ionized oxygen should produce an O~{\small VIII} radiative
recombination continuum at around 14 \AA\, (14.2 \AA\, for plasma at
rest and 14.4 \AA\, at $z=0.01$).  The RGS spectrum shows some
residuals at this location. Including a recombination emission edge
at 14.4 \AA~ with a 50 eV width in the fit gives a normalization of
(6.1$\pm$3.1)$\times$10$^{-4}$ photons cm$^{-2}$ s$^{-1}$. The
addition of this component is statistically significant, although it
is quite broad and could also result from uncertainties in the
spectral continuum model. The RGS spectra were also rebinned using
different criteria, but no other spectral features could be
identified.

\begin{figure}
\psfig{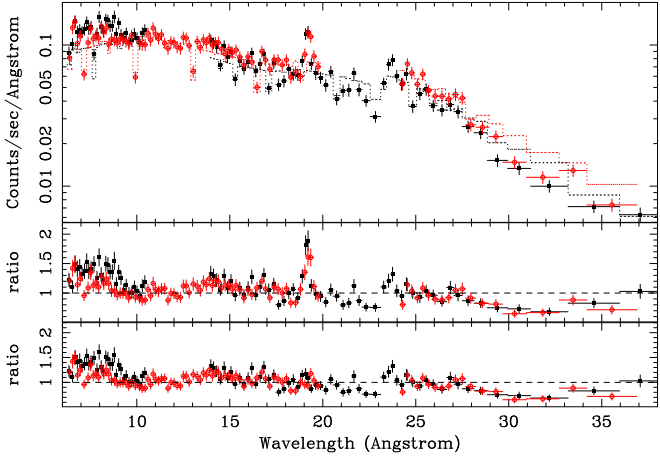} \caption{RGS1 (filled
squares) and RGS2 (open circles) spectra of \uu. The residuals in
the middle panel are relative to the model derived from the PN data
analysis (see table 1) multiplied by a factor 0.8 to account for the
discrepancies in flux estimates between the PN and RGS. The lower
panel shows the residuals relative to the same model plus an
emission line at 19.19 \AA. For clarity, the data have been rebinned
to a bin size larger than the RGS energy resolution. The data gaps
at 10.5--14 \AA~ for RGS1 and 20 -- 24 \AA~ for RGS2 correspond to
the CCDs lost due to electronic failure in the early phase of the
mission. \label{fig2}}
\end{figure}

Due to the poorer energy resolution of the PN camera, no line at
19.19 \AA~ is significantly detected in the PN data. However, if we
fix the energy and the width of the line to the values derived from
the RGS data, the upper limit to the line equivalent width is 12 eV,
which is consistent with the RGS results.

\section{Discussion}

The high sensitivity of \xmm at low energies and the high spectral
resolution of the RGS instrument have revealed new features in the
X-ray spectrum of the unusual interacting binary \uu. Although the
PN spectrum above 1 keV is consistent with the rather variable
spectra seen in previous observations (Masetti et al. 2002), we
clearly detect both a soft excess and an emission line at 19.2 \AA.

The best fit to the
soft excess was found by adding
a broad Gaussian line component to the high-energy continuum model
(blackbody + Comptonization).  But the interpretation of this feature
as an emission line is problematic.
No strong emission lines are expected at the wavelength of the soft
excess for either $z=$0.01 or $z=$0.  In addition, both the intensity
and broadening of this line would be much larger than for the line at 19.2
\AA.

As mentioned above, we have also discovered an emission line at 19.2 \AA.
Since only very low emissivity lines are
consistent with the observed wavelength of $\lambda$ = 19.19$^{+0.05}_{-0.09}$
\AA, the line may be O~{\small VIII} at
redshift $\simeq$0.01.
For most high-emissivity lines close to this wavelength, a much larger
redshift/blueshift would be required.  For example, among the other
strong lines expected in the spectrum of an accreting X-ray binary,
the O~{\small VII} and Ne~{\small IX} triplets are the nearest
candidates for the line identification, at longer and shorter
wavelengths, respectively. For O~{\small VII}, the line would be
blue-shifted by a factor $>0.1$, which, in case of Doppler shift,
requires that the emitting plasma is moving towards us at more than
10\% of the speed of light. On 2002 July 29 radio observations were
performed to look for a possible jet, but no source was detected with
an upper limit of 1.0$\pm0.7$ mJy at 15 GHz (G. Pooley 2002, private
communication).

On the other hand, the identification of the line at 19.2 \AA\, with
the Ne~{\small IX} triplet would imply redshifts of 0.40, 0.41, and
0.42 for the forbidden, intercombination and resonance lines,
respectively. These values could be interpreted as gravitational
redshift from close to the surface of a neutron star, as they are
consistent with most of the equations of state for neutron stars
composed of normal nuclear matter, and they are just slightly larger
than the redshift $z$=0.35 found by Cottam et. al. (2002) in the
spectral analysis of X-ray bursts from EXO 0748--676.  However, due
to the lack of identifications of other spectral features at $z
\sim$ 0.4 or indication for overabundance of neon, we consider
O~{\small VIII} to be a better candidate for the observed emission
line.

The relatively small redshift of the O~{\small VIII} line ($z \simeq$
0.01) can be interpreted in several ways, and we briefly discuss two
possible scenarios.  A value of $z$ = 0.008--0.014 corresponds to the
gravitational redshift of a photon emitted at a distance of 35--60
Schwarzschild radii from a compact object: this interpretation would
exclude the possibility that \uu is a white dwarf and would correspond
to a distance of 1.5--2.6$\times$10$^{7}$ cm from a 1.4 $\msole$
neutron star.
In a photoionized plasma, assuming that most of the oxygen  is H-like,
the measured O~{\small VIII} line luminosity of $\simeq$10$^{40}$
photons s$^{-1}$ gives an emission measure of $EM \simeq 6 \times
10^{53}$ cm$^{-3}$ (taking an abundance of $\simeq$1.6$\times
10^{-3}$, as derived from the oxygen edge fit and assuming that the
overabundance is intrinsic to the X--ray source). Assuming a
spherically symmetric geometry, if the O~{\small VIII} line is emitted
at $\simeq 2\times$10$^{7}$ cm from the neutron star, we can estimate
a density of $n \simeq 4
\times 10^{15}$ cm$^{-3}$. For the measured continuum luminosity of
10$^{34}$ erg s$^{-1}$, the ionization parameter is $\xi
=$L$_{cont}/($n R$^2)\simeq$5000 erg cm s$^{-1}$. For such a high
ionization parameter, most of the oxygen should be fully ionized and
therefore we would not expect to see a prominent O~{\small VIII}
emission line. O~{\small VIII} dominates the ionization states of
oxygen for $\xi \simeq 100$ erg cm s$^{-1}$, which means that either
we have used an oversimplified model (assuming, for example, that
the
line emission region is symmetric and homogeneous) or that the
O~{\small VIII} line is emitted at larger distance from the central
X-ray source and therefore that the redshift is not solely
gravitational.

In an alternative scenario, the same redshift can be produced by
Doppler effects if the emitting plasma is moving away from us at a
speed of 2000--4000 km/s. This velocity is about two orders of
magnitude larger than the wind velocity of an M-type giant (Reimers
1977), as well as the proper motion and possible orbital velocity of
the binary system (Galloway, Sokoloski \& Kenyon 2002). Although the
high luminosity, hard X-ray spectrum, and rapid X-ray variability of
\uu make it rather unlikely, the present data do not rule out the
possibility that the accreting object is a white dwarf. The
detection of Doppler-shifted lines of highly ionized elements has
been reported for some supersoft sources and interpreted as the
signature of collimated outflows ("jets") coming from the accreting
white dwarf (e.g., Cowley et al. 1998).  Furthermore, the jet
velocity is typically similar to the value we derive for the
redshifted O~{\small VIII} line.  In our case, however, no
corresponding blueshifted line is detected, which implies either a
unipolar jet or some special geometry to obscure the approaching
jet.

For a neutron star with a $10^{12}$ gauss magnetic field and a
luminosity of $10^{34}$ erg s$^{-1}$, the magnetospheric radius is
$\simeq 3\times$ 10$^{9}$ cm (Hayakawa 1985).
Although X-ray pulsations have not been found from \uu,
the fact that it is probably viewed close to face-on (Galloway et
al. 2002) means that pulsations would be undetectable even if it is an
accreting neutron star with high magnetic field as the line of
sight is almost aligned with the neutron star spin axis.
At a distance of $\simeq 3\times$ 10$^{9}$ cm from a 1.4 $\msole$
neutron star, a free particle would move radially towards the neutron
star with a velocity of $\simeq 3400$ km/s. Therefore, the redshift we
measure could originate from
close to the magnetospheric radius, where the plasma density is
increased by the funneling effect produced by the magnetic field
lines.  Repeating the estimates of the ionization parameter described
above, at a distance of 3$\times$ 10$^{9}$ cm the corresponding
density is $n \simeq 2 \times 10^{12}$ cm$^{-3}$ and $\xi \simeq
500$. These two values are fairly consistent with the presence of an
O~{\small VIII} line.

Redshifted emission lines have also been reported in Cygnus X-3.
Paerels et al. (2000) reported a redshift corresponding to 750--800
km/s for all the emission lines detected in a \cha HETGS observation
of Cygnus X-3. In two more recent \cha observations, a slightly
smaller redshift of 270--550 km/s, in addition to a Doppler modulation
of about 150 km/s, was found (Stark \& Saia 2003). The second effect
was interpreted as being due to the binary orbital motion, but no
convincing explanation was given for the global redshift. Although
the X-ray luminosities and observed redshifts of
\uu and Cygnus X-3 are rather different, a common mechanism could
produce the line redshifts.
For both \uu and Cygnus X-3, a low orbital inclination has been
proposed (Galloway et al. 2002; Mioduszewski et al. 2001).  Thus,
Doppler shifts due to the bulk velocity of matter
receding from us as it falls onto the compact object
could be the origin of the redshifted lines.

\begin{acknowledgements}
Based on observations obtained with XMM-Newton, an ESA science
 mission  with  instruments and contributions directly funded by ESA Member States and NASA.
This work was partially supported by the Netherlands Organization for
Scientific Research (NWO).
This work was supported in part by the NASA Long Term
Space Astrophysics program under grant NAG 5-9184.
JMM and JLS gratefully acknowledge support from the NSF.
\end{acknowledgements}

\end{document}